\documentclass[a4paper,fleqn,usenatbib]{mnras}

\usepackage[T1]{fontenc}
\usepackage{ae,aecompl}

\usepackage{graphicx}	
\usepackage{amsmath}	
\usepackage{amssymb}	
\usepackage{lscape}
\usepackage{natbib}
\usepackage[varg]{txfonts}
\usepackage{url}
\usepackage{xspace}
\usepackage{longtable}
\usepackage{pdflscape}	
\bibpunct{(}{)}{;}{a}{}{,}
\pdfoutput=1

\newcommand{\sao}{\mbox{SAO\,244567}\xspace}

\newcounter{Rco}
\newcommand{\Ionst}[1]{\setcounter{Rco}{#1}\Roman{Rco}}
\newcommand{\Ion}[2]{\mbox{#1\,{\scriptsize\Ionst{#2}}}}
\newcommand{\Ionw}[3]{\mbox{#1\,{\scriptsize\Ionst{#2}}~$\lambda\,#3$\,\AA}}

\newcommand{\Ionww}[3]{\mbox{#1\,{\scriptsize\Ionst{#2}}~$\lambda\lambda\,#3$\,\AA}}

\newcommand{\logg}{\mbox{$\log g$}\xspace}

\newcommand{\se}[1]{\mbox{Sect.\,\ref{#1}}}

\newcommand{\sla}{\raisebox{-0.10em}{$\stackrel{<}{{\mbox{\tiny $\sim$}}}$}}

\newcommand{\Teff}{\mbox{$T_\mathrm{eff}$}\xspace}

\newcommand{\Lsol}{$L_\odot$}
\newcommand{\Msol}{$M_\odot$}

\newcommand{\Mdot}{$\dot{M}$}
\newcommand{\pa}{\mbox{Paper\,I}}

\title[\sao is now returning towards the AGB]{Breaking news from the HST: The central star of the Stingray Nebula is now returning towards the AGB}
\author[Reindl et al.]{
Nicole Reindl,$^{1,2}$\thanks{E-mail: nr152@le.ac.uk}
T. Rauch,$^{2}$
M. M. Miller Bertolami,$^{3}$
H. Todt,$^{4}$
K. Werner$^{2}$
\\
$^{1}$Department of Physics and Astronomy, University of Leicester, University Road, Leicester LE1 7RH, UK\\
$^{2}$Institute for Astronomy and Astrophysics, Kepler Center for Astro and Particle Physics, Eberhard Karls University, Sand 1, 72076 T\"ubingen, Germany\\
$^{3}$Instituto de Astrof\'{i}sica de La Plata, UNLP-CONICET, La Plata, Buenos Aires, 1900, Argentina\\
$^{4}$Institute for Physics and Astronomy, University of Potsdam, Karl-Liebknecht-Str. 24/25, 14476 Potsdam, Germany
}
\date{Accepted XXX. Received YYY; in original form ZZZ}

\pubyear{2016}

\begin{document}
\label{firstpage}
\pagerange{\pageref{firstpage}--\pageref{lastpage}}
\maketitle

\begin{abstract}
\sao is a rare example of a star that allows us to witness stellar evolution in real time. 
Between 1971 and 1990 it changed from a B-type star into the hot central star of the Stingray Nebula. 
This observed rapid heating has been a mystery for decades, since it is in strong contradiction with the low 
mass of the star and canonical post-asymptotic giant branch (AGB) evolution. We speculated that \sao might 
have suffered from a late thermal pulse (LTP) and obtained new observations with HST/COS to follow the evolution of 
the surface properties of \sao and to verify the LTP hypothesis. Our non-LTE spectral analysis reveals that the star 
cooled significantly since 2002 and that its envelope is now expanding. Therefore, we conclude that \sao is currently 
on its way back towards the AGB, which strongly supports the LTP hypothesis. A comparison with 
state-of-the-art LTP evolutionary calculations shows that these models cannot fully reproduce the evolution of all surface parameters 
simultaneously, pointing out possible shortcomings of stellar evolution models. Thereby, \sao keeps on challenging 
stellar evolution theory and we highly encourage further investigations. 
\end{abstract}

\begin{keywords}
stars: AGB and post-AGB -- stars: atmospheres -- stars: evolution
\end{keywords}



\section{Introduction}

Revealing the evolution of stars is a complex task, since their evolutionary history can usually only be 
reconstructed indirectly due to the long evolutionary times scales. 
In a few cases, however, the evolution of the surface properties of a star occurs on a time-scale shorter 
than a human lifetime. Such events allow us to witness stellar evolution in real time and provide a unique 
way to gain a direct knowledge of stellar evolution.\\
\sao, the ionization source of the Stingray Nebula (\mbox{PN\,G331.3$-$12.1}), is one such 
rare example of an unusually fast evolving star. Rapid changes of its observable properties were at first noticed by 
\cite{partha1993, partha1995}. Based on the optical spectrum of \sao obtained in 1971 and its \textit{UBV} coulors, they 
concluded that \sao was a B-type star with an effective temperature of \Teff\ $\approx 21$\,kK at that time. 
However, the optical spectra from 1990 and 1992 as well as IUE spectra from 1992 on display many nebular emission 
lines, which let them conclude that \sao has ionized its surrounding nebula within only two decades. 
In \citeauthor{Reindletal2014a} (\citeyear{Reindletal2014a}, \pa), we presented first quantitative spectral analyses 
of all available spectra taken from 1988 to 2006 with various space-based telescopes. We found that the central 
star had steadily increased its \Teff\ from 38\,kK in 1988 to a peak value of 60\,kK in 2002. During the same time, 
the star was contracting, as concluded from an increase in surface gravity from $\log$(g\,/\,cm\,s$^{-2}$)\,=\,4.8 to 
6.0 and a drop in luminosity. Simultaneously, the mass-loss rate declined from 
$\log$(\Mdot\,/\,\Msol\,yr$^{-1}$)\,=\,$-9.0$ to $-11.6$ and the terminal wind velocity increased from 
$v_\infty$ = $1800$ to $2800$\,km\,s$^{-1}$.\\
The existence of the planetary nebula (PN) suggests that \sao is likely a post-AGB star because -- in general -- only 
such a star is expected to eject a PN. On the other hand, its low mass is in strong contradiction with canonical 
post-AGB evolution \citep{partha1995, Bobrowsky1998}. A comparison of the position of \sao in the log \Teff\ -- \logg\ 
plane with latest post-AGB stellar evolutionary calculations \citep{MillerBertolami2016} indicates a mass below 
$0.53$\,\Msol, while the rapid heating rate (d\Teff\,/\,d$t$) would require a central star mass of 
$0.7$\,\Msol.\\
Under certain circumstances, however, the classical picture of the evolution of post-AGB stars is altered by the 
occurrence of a late He-shell flash. These can occur either during the blue-ward excursion of the post-AGB star 
(late thermal pulse, LTP), or during its early white dwarf cooling phase (very late thermal pulse, VLTP). The 
release of nuclear energy by the flashing He-shell forces the already very compact star to expand back to giant 
dimensions -- the so-called born-again scenario. This scenario was first explored in detail by 
\cite{Schoenberner1979} and by \cite{Iben1983} and is of great importance for the explanation of H-deficient post-AGB 
stars, which make up about a quarter of the post-AGB stars. The evolutionary time-scales of stars, which are 
considered to have undergone such a late He-shell flash, are very short (decades) and thus, the detection and the 
repeated observation of such quickly evolving objects make it possible to record the temporal evolution of their 
stellar parameters and to establish constraints for stellar evolution theory. Well known examples for 
``born-again'' stars are \mbox{V605\,Aql} (e.g., \citealt{Clayton2006}), \mbox{V4334\,Sgr} (Sakurai's object, e.g. 
\citealt{Hajduk2005}) and FG\,Sge (e.g., \citealt{Jeffery2006}). \mbox{V605\,Aql} and \mbox{V4334\,Sgr} are 
considered to have undergone a VLTP, which produces an H-free stellar surface already during the flash. FG\,Sge, on the 
other hand, had a slower cooling rate and turned H-deficient only when it had returned back to the AGB. Therefore FG\,Sge 
must instead have experienced an LTP (see \citealt{Schoenberner2008} for a review about these objects).\\
In \pa, we speculated that \sao might also have experienced a late He-shell flash shortly after leaving the AGB.
This scenario predicts that \sao would eventually become a cool supergiant, i.e., that the star 
will become cooler and expand. To follow the evolution of the surface properties of \sao and to verify the LTP hypothesis, 
we performed further observations with the Cosmic Origins Spectrograph (COS) on the Hubble Space Telescope (HST).

\section{Observations}
\label{sect:observations}

We obtained HST/COS spectra using the PSA aperture (total exposure time 13.4\,ks, proposal ID 13708). 
The observations were performed on the 2015 August 9, more than nine years after the star 
was previously observed with an ultraviolet (UV) telescope. Far-UV medium and low 
resolution spectra were obtained using gratings G130M and G140L at central wavelength settings 1222\,\AA\ and 1280\,\AA, 
respectively. In addition, we acquired near-UV low resolution spectra with the G230L grating at central 
wavelength settings 2635\,\AA\ and 3360\,\AA. The latter are unfortunately blurred by the strong \mbox{$[$\Ion{C}{2}$]$} 
$\lambda\,2326$\,\AA\ (nebular) and $[$\Ion{O}{2}$]\,\,\lambda\,2471$\,\AA\ (nebular and geocoronal) emission lines and, 
hence, provide neither reliable fluxes nor line profiles. This is due to the very compact PN 
($\tiny{\diameter}\approx 2$\arcsec) and the much larger aperture used for COS ($2$\farcs$5$) observations 
compared to the apertures used for the former STIS ($52$\arcsec$\times 0$ \farcs$05$) and FOS ($0$\farcs$3$) observations. 
Comparing the new HST/COS observation to the FUSE observations in 2002 and 2006, we found that the flux decreased slightly 
by a factor of 1.6 and 1.4, respectively. 

\section{Spectral analysis}
\label{sect:analysis}

The spectral analysis was carried out analogously to \pa\ by modelling both the photospheric and 
the interstellar line-absorption spectrum. Again, we employ the OWENS program to model the ISM line absorption. The 
interstellar reddening was derived as outlined in \pa. Using the LMC reddening law of \cite{howarth1983}, we 
found \textit{E(B-V)}\,=\,$0.18\pm0.02$. For the quantitative spectral analysis we used the T{\"u}bingen non-LTE 
Model-Atmosphere Package (TMAP, \citealt{werneretal2003, tmap2012, rauchdeetjen2003}), which allows the computation 
of fully metal-line blanketed model atmospheres in radiative and hydrostatic equilibrium (\se{subsect:TMAP}). In 
order to constrain the presence of a possible weak stellar wind we employed PoWR 
\citep[Potsdam Wolf-Rayet model-atmosphere code,][\se{subsect:PoWR}]{graefeneretal2002, hamannetal2003, hamannetal2004}.
\begin{figure}
\includegraphics[width=\columnwidth]{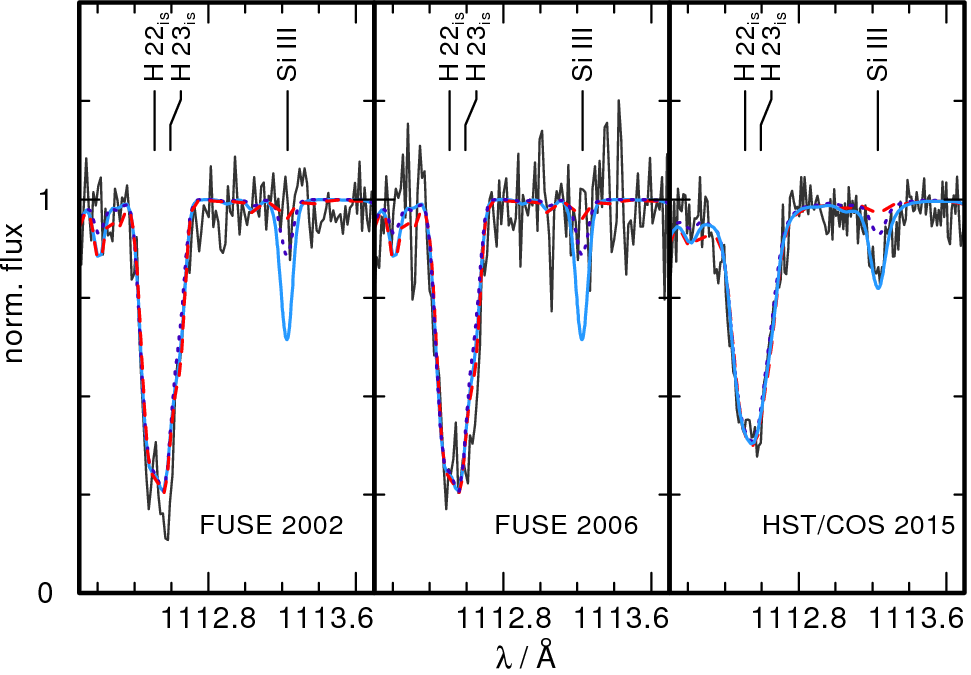}
\vspace*{-5mm}
\caption{Emergence of Si III $\lambda$ 1113.2\,\AA\ in the HST/COS observation taken in 2015 (right). This 
line was not visible in the FUSE observation in 2002/2006 (left/middle) when the star was about 10/5\,kK hotter. 
The dotted / dashed line corresponds to TMAP models with \Teff\,=\,$60/55$\,kK (best fit for 2002/2006), while the 
solid line represents a model with \Teff = $50$\,kK (best fit for 2015). Note that the differences in the line 
profiles are caused by the instruments' line spread functions.}
\label{TeffSi}
\end{figure}

\subsection{TMAP model atmosphere analysis}
\label{subsect:TMAP}

As a first step we compared the new observations with the best-fitting model atmosphere for 2006 (year of the last UV observations, see \pa), 
which included the elements H, He, C, N, O, Si, P, S, Fe, and Ni to 
identify several interstellar and photospheric spectral lines. Thanks to the better S/N of the COS G130M spectrum and 
the extent towards larger wavelengths we were able to identify numerous lines of Fe, and for the first time also Ni. 
By comparison with the grid calculated in \pa, we found $\mathrm{Ni}$\,=\,$6.8\times 10^{-5}$ (mass fraction, 
which is solar according to \citealt{Scottetal2015b}). Surprisingly, we discovered the emergence of the 
\Ionw{Si}{3}{1113.2} line in the new HST/COS spectra (Fig.~\ref{TeffSi}). This line was not visible in the previous 
FUSE observations, which cover this wavelength range as well and have about the same resolving power 
($R \approx 20\,000$).\\ 
Some lines, however remained unidentified. In order to investigate the origin of those lines we proceeded as follows.
To check for the presence of lines from iron-group elements, besides Fe and Ni, we calculated TMAP models including H and He, 
plus Ca, Sc, Ti, V, Cr, or Mn (at abundances $0.1, 1 ,10, 100\,\times$ solar, according to \citealt{Scottetal2015a, Scottetal2015b}), 
respectively. These were then compared to the G130M observation in order to derive the abundances or at least upper limits. 
The model atoms were calculated via the T{\"u}bingen iron-group opacity interface TIRO \citep{ringatPhD2013}. 
No lines of Sc are predicted in the G130M wavelength range and the strongest line of calcium (\Ionw{Ca}{4}{1096.7}) is blended 
by interstellar lines of H\,$_{\mathrm{23}}$ and \Ion{Fe}{2}. Therefore no upper limits are given for these elements. 
The strongest predicted lines of titanium, namely \Ionww{Ti}{4}{1183.6, 1195.2} and \Ionww{Ti}{5}{1153.3, 1168.1, 1230.4} are not 
visible in the observations. We derive an upper limit of $\mathrm{Ti}\,\sla\,3.0\times 10^{-4}$ ($100\times$ solar). Also the 
strongest predicted lines of vanadium (\Ionww{V}{5}{1142.7, 1157.6}), as well as manganese (\Ionww{Mn}{6}{1285.1, 1333.9} and 
\Ionw{Mn}{5}{1359.2}) are not detected. We derive upper limits of $\mathrm{V}\,\sla\,2.9\times 10^{-7}$ 
(solar) and $\mathrm{Mn}\,\sla\,1.0\times 10^{-5}$ (solar), respectively. For the first time we discovered 
\Ionww{Cr}{5}{1121.1, 1127.6} and derived Cr\,=\,$1.6\times 10^{-5}$ (solar).\\
\begin{figure}
  \includegraphics[width=\columnwidth]{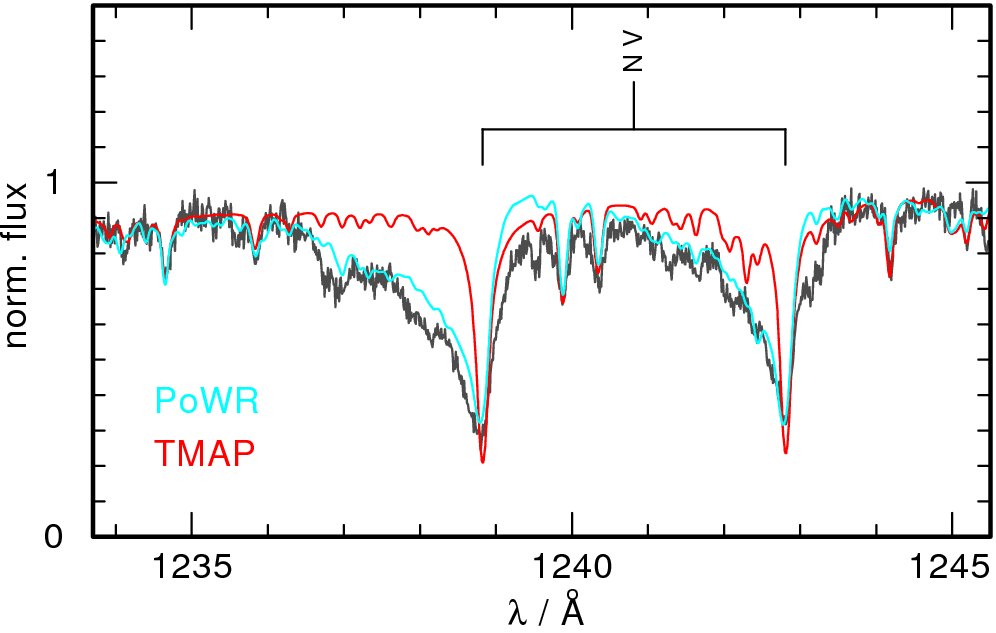}
\vspace*{-5mm}
  \caption{The \Ion{N}{5} resonance doublet as observed in 2015 with HST/COS (grey). It is not possible to reproduce the 
  asymmetrical line profiles with the static TMAP model (red, \Teff\,=\,$50$\,kK, \logg\,=\,5.5). In the PoWR model (blue, $10^{-12}$\,\Msol\,yr$^{-1}$, 
  \mbox{$v_\infty$\,=\,$500$\,km\,s$^{-1}$}, however, the blue-shifted absorption wings are reproduced well with \mbox{$v_\infty$\,=\,$500$\,km\,s$^{-1}$}. 
  The emission peak of the P-Cygni profiles vanishes when \Mdot\ falls below $10^{-12}$\,\Msol\,yr$^{-1}$.}
\vspace*{-5mm}
  \label{fig:wind}
\end{figure}
As a next step, we increased the numbers of non-LTE levels for O, Si, P, and S for our model atmosphere calculations. We then 
calculated a model grid spanning \Teff\,=\,$45\,-\,60$\,kK (step size $2.5$\,K) and including the elements H, He, C, N, O, 
Si, P, S, Cr, Fe, and Ni (at abundances as derived above and in \pa). 
Because the near-UV COS spectrum, which we originally obtained to derive \logg\ via the \Ion{He}{2} lines 
($n\,\rightarrow\,n'$\,=\,$3\,\rightarrow\,6, ..., 14$), is not suited for the spectral analysis (see \se{sect:observations}), 
we assumed initially \logg = $6.0$, i.e., that the surface gravity has not changed significantly. Note that the surface gravity 
is further investigated below, but the next step only focused on the \Teff\ determination by comparing our grid to 
the new observations.\\
We found that \Ionw{Si}{3}{1113.2} was not visible in the FUSE observation in 2002, because the star was about $10$\,kK 
hotter in 2002 compared to 2015 where we find the best fit at \mbox{\Teff\,=\,$50$\,kK}. The strengths of \Ionww{Si}{4}{1128.3, 1154.6} 
do not show a strong dependence on \Teff\ in the temperature region studied, thus, the Si abundance derived in \pa\ is 
reliable. This value is also confirmed by evaluating additionally the ionization equilibria of 
\Ion{C}{3}\,/\,\Ion{C}{4}, 
\Ion{N}{3}\,/\,\Ion{N}{4}, 
\Ion{O}{3}\,/\,\Ion{O}{4}\,/\,\Ion{O}{5},
\Ion{S}{4}\,/\,\Ion{S}{5}\,/\,\Ion{S}{6}, 
\Ion{Fe}{5}\,/\,\Ion{Fe}{6}, and 
\Ion{Ni}{4}\,/\,\Ion{Ni}{5}. 
The good S/N allows the error on \Teff to be reduced to $\pm 2.5$\,kK. For \Teff $\ge 47.5$\,kK the lines of \Ion{N}{3}, \Ion{O}{3}, 
\Ion{Si}{3}, and \Ion{S}{4} are too strong in the models, whereat \Ionw{S}{6}{1117.8} becomes too weak. In contrast, for 
\Teff $\ge 52.5$\,kK the lines of \Ion{C}{3}, \Ion{N}{3}, \Ion{O}{3}, \Ion{Si}{3}, and \Ion{S}{4} are too weak, while the lines 
of \Ion{N}{4}, \Ion{S}{6}, and \Ion{Fe}{6} are too strong in our models. The line profiles of 
\Ion{C}{4}, \Ion{Si}{4}, \Ion{S}{5}, \Ion{Ni}{4} and \Ion{Ni}{5} show only mild changes in the range \Teff\ = $50\pm5$\,kK. The same 
holds for \Ionww{O}{4}{1338.6-1343.5} while the line strength of \Ionww{O}{4}{1164.3, 1164.5} increases significantly for 
\mbox{\Teff\ $\geq 50$\,kK}. The latter two lines were not included in the model of \pa, and therefore not identified. However, 
these lines are clearly detected in the FUSE observations from 2002 and 2006.\\
Finally, we investigated a possible impact of \logg\ on the derived \Teff\ by calculating a second model grid with 
\mbox{\logg = 5.5}. We found that the metal lines hardly differ from the \mbox{\logg = 6.0} model and that the ionization 
equilibria are also reproduced best at \mbox{\Teff = $50\pm2.5$\,kK}. Furthermore, we find that \Ionw{He}{2}{1640.4} (Fig.~\ref{fig:G140L}) 
is best reproduced at \mbox{\logg = $5.5\pm0.5$}. For larger/lower values of \logg, the line wings of our models become too broad/narrow 
compared to the observations.

\subsection{PoWR model atmosphere analysis}
\label{subsect:PoWR}

Conclusions about the properties of the stellar wind can be drawn by studying prominent wind lines located in the far-UV. 
The S/N of the G140L spectrum is poor for $\lambda<1050$\,\AA\ and cannot be improved sufficiently by data binning, which 
prevented investigations of the \Ion{O}{6} resonance doublet. The \Ion{C}{4} resonance doublet, also recorded with the G140L grating, does 
not show any signature of a stellar wind (Fig.~\ref{fig:G140L}). \Ionww{N}{5}{1238.8, 1242.8}, however, show asymmetrical line profiles, 
i.e., they show blue-shifted absorption wings while an emission, which would clearly point to a P-Cygni profile, cannot be detected 
(Fig.~\ref{fig:wind}). To investigate the origin of the asymmetry, we calculated non-LTE models for expanding model atmospheres 
with PoWR. We assumed \mbox{$M$ = $0.5$\,\Msol}, \mbox{$\log$($L$\,/\,\Lsol)\,=\,$2.5$}, \mbox{\Teff\,=\,$50$\,kK} and elemental 
abundances as found in the TMAP analysis. 
In the absence of diagnostic lines, e.g., recombination emission lines vs. electron scattering line wings, we assume a smooth 
wind for the calculation of the population numbers. In the formal integral we introduce optically thick inhomogeneities (see \cite{Oskinovaetal2007}) 
to achieve a better fit of the \Ion{N}{5} resonance doublet line cores, which otherwise appear too strong in the PoWR model.
As the line opacity in the wind is already very low, the absorption wings of the P-Cygni profiles are less affected by the porosity effect.\\
We determine an upper limit for the mass loss rate of \Mdot\,=\,$10^{-12}$\,\Msol\,yr$^{-1}$. Below this value the emission peaks of the 
\Ion{N}{5} lines vanish (Fig.~\ref{fig:wind}). The terminal wind velocity was measured from the blue edge of the 
absorption component of \Ionw{N}{5}{1238.8}. We found $v_\infty$\,=\,$500$\,km\,s$^{-1}$, which is significantly less than what was 
found for the previous years ($v_\infty$ = $2800$\,km\,s$^{-1}$ in the year 2006). This drastic change implies that also the 
surface gravity must have decreased by half an order of magnitude presuming $v_\infty  \propto v_{\mathrm{esc}} \propto g$. 
Therefore a change from \mbox{\logg = 6.0} in 2006 to \mbox{\logg\,=\,5.5} in 2015 is perfectly supported.\\
Note that the PoWR model (with  \Mdot\,=\,$10^{-12}$\,\Msol\,yr$^{-1}$, $v_\infty$\,=\,$500$\,km\,s$^{-1}$) does not predict P-Cygni profiles for 
other prominent wind lines (e.g., the \Ion{O}{6} and \Ion{C}{4} resonance doublets). This is in line with previous IUE observations in 
from 1994 to 1996 in which also only the resonance doublet of \Ion{N}{5} exhibited P-Cygni profiles and \sao had \mbox{\Teff\,=\,$50$\,kK} 
as well.

\section{Results and discussion} 
\label{sect:discussion}

\begin{figure}
  \includegraphics[width=\columnwidth]{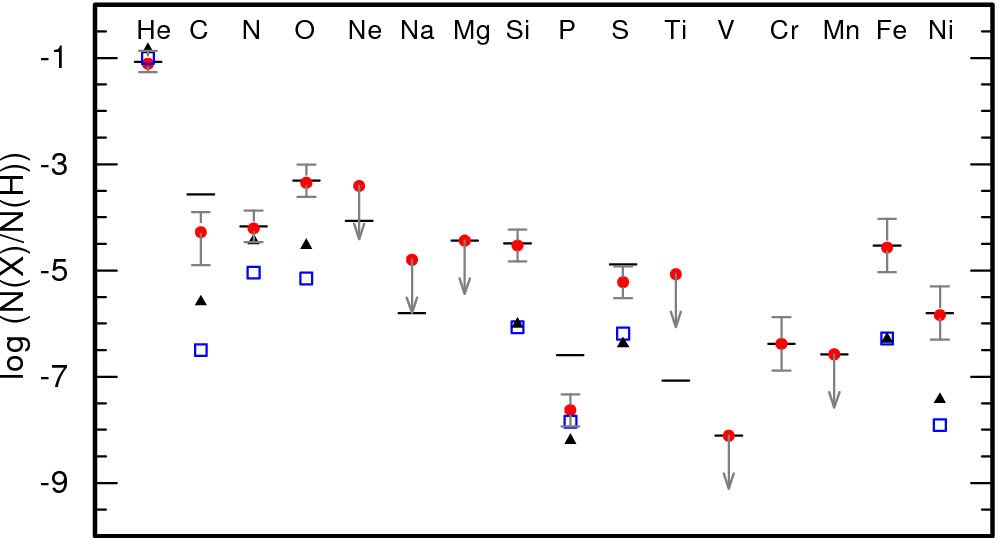}
\vspace*{-5mm}
  \caption{Photospheric abundances of \sao relative to H (number ratios, red circles). Solar abundance values are indicated by black lines. The abundances of two other lower metallicity low-mass post-AGB stars (ROB\,162, blue squares, Chayer in prep., and vZ\,1128, black triangles, (\citealt{Chayeretal2015}) are also shown.}
  \label{fig:abund}
\end{figure}

\begin{figure}
\includegraphics[width=\columnwidth]{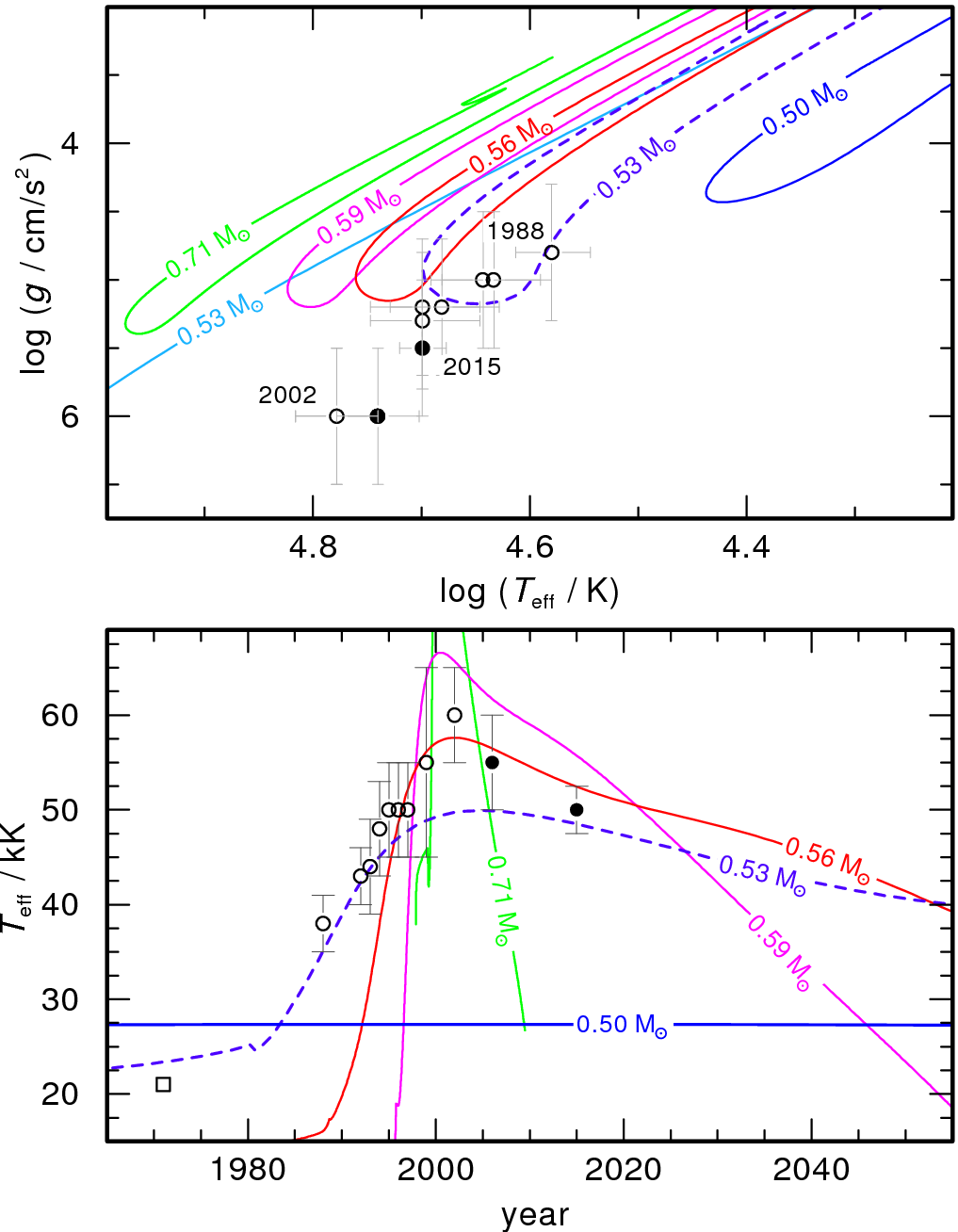}
\vspace*{-5mm}
\caption{Upper panel: Evolution of \sao (blue/red-wards evolution is indicated by open/filled circles) in the log \Teff\ -- \logg\ compared with different post-AGB evolutionary tracks (labelled with stellar masses, see text for details). The light blue line represents a canonical H-burning post-AGB, while the other models suffer from a LTP. Models that were calculated with Z\,=\,0.001 are shown as solid lines, while the dashed line represents a model with Z\,=\,0.02. Lower panel: Temporal evolution of \Teff\ as observed and as predicted by the LTP models during the \Teff\ reversal. The open square indicates the \Teff\ estimate when \sao was still a B-type star \citep{partha1995}.}
\label{fig:LTP} 
\end{figure}

The new HST/COS spectra allows us to identify lines of Cr and Ni and to measure their abundances for the first time. We 
confirm the abundance values of all the other elements derived in \pa, i.e., no hint of a change in the chemical 
abundances is found. Figure~\ref{fig:abund} shows the elemental abundances of \sao as derived in this work and \pa. 
The largely solar surface composition is in line with the thin disk nature of \sao as concluded from its space velocities. 
These were calculated using $v_{\mathrm{rad}}$=$14\pm 12$\,km/s (measured from the G130M spectrum), proper motions 
and statistical parallax from \cite{Fresneauetal2007}. We found ($U, V, W$)\,=\,($32\pm18, 27\pm13, 4\pm6$)\,km/s, which is 
typical for thin disk stars \citep{Kordopatis2011}. We note, that the height of \sao above the Galactic plane 
($z$\,=\,$150$\,pc assuming $d$\,=\,$826$\,pc from \citealt{Fresneauetal2007}) also lies within the scale height of the thin 
disk.\\
The solar abundances of He and N excluded the occurrence of the second dredge-up and hot bottom burning. This agrees with 
the fact that \sao must be a low mass star since the second dredge-up and hot bottom burning occurs only in stars 
with initial mass $M>3-4$\Msol\ \citep{GarziaHernandez2013}. In addition the sub-solar C abundance indicates that \sao has 
not experienced a third dredge-up (\pa), which implies that in case \sao was once a AGB star, it must have had a low 
initial mass ($M\lesssim1.5$\Msol, \citealt{Cristalloetal2015, GirardiMarigo2007}). 
Two well studied low-mass post-AGB stars that also avoided the third dredge-up are \mbox{vZ\,1128} 
\citep{Chayeretal2015} and \mbox{ROB\,162} (Chayer et al. in 
prep\footnote{\url{http://www-astro.physics.ox.ac.uk/~aelg/SDOB7/proceedings/Chayer_rob162_proceedings.pdf}}.), 
whose abundances are shown in Fig.~\ref{fig:abund}, too. They display very similar abundance patterns to \sao, although the 
metal abundances are systematically lower by about one order of magnitude. This can, however, be simply understood in terms 
of the lower initial metallicities of \mbox{vZ\,1128} and \mbox{ROB\,162}, since they are found in globular clusters 
(\mbox{NGC\,5272} and \mbox{NGC\,6397}, respectively). \citeauthor{Chayeretal2015} (2015, in prep.) demonstrated that their abundances 
coincide with the abundances of red giant branch (RGB) and horizontal branch (HB) stars in these clusters, i.e., that 
the abundances of these stars have not changed since the RGB. 
The positions of \mbox{vZ\,1128} and \mbox{ROB\,162} in the $\log$ \Teff\ -- \logg\ plane agree with a 
$0.53$\,\Msol\ post-AGB evolutionary track of \citeauthor{MillerBertolami2016} (2016, see Fig.\,5 in 
\citealt{Reindletal2016}), while \sao seems to have an even lower mass (Fig~\ref{fig:LTP}). In \pa, we 
discussed the possibilities of \sao being a post-RGB or post-extreme HB star and we will reconsider these later 
on. From the star's surface abundances, kinematics, height above the Galactic plane and the position in the 
$\log$ \Teff\ -- \logg\ plane, we can, for now, maintain that \sao is a low-mass, population I 
star.\\
The most striking result that we obtained from the new HST observations is the fact that \sao cooled significantly since 2002. 
While in 2002 the star had \Teff = $60\pm5$\,kK, we find that in 2015 the ionization equilibria of several species are best 
reproduced at \Teff\,=\,$50\pm2.5$\,kK. Moreover, we found evidence, that \sao must have been expanding since 2006. This is 
concluded from the decrease of the terminal wind velocity by about factor of five since 2006 ($v_\infty$\,=\,$500$\,km\,s$^{-1}$ 
in 2015, $v_\infty$\,=\,$2800$\,km\,s$^{-1}$ in 2006). This is in line with a decrease of the surface gravity from 
\mbox{\logg\,=\,$6.0$} to \mbox{\logg\,=\,$5.5$}. Finally, we found that the mass loss rate of \sao has further declined 
(\Mdot $\leq 10^{-12}$\,\Msol\,yr$^{-1}$, compared to \Mdot =$10^{-11.6}$\,\Msol\,yr$^{-1}$ in 2006).\\
From the initial rapid heating and contraction of \sao (heating rate from 1988 to 2002: $1570 \pm 570$\,K/yr), which is now 
followed by a rapid cooling and expansion of the star, we conclude that \sao has indeed suffered a late He-shell flash and is now on its way 
back towards the AGB. The cooling rate of \sao ($770 \pm 580$\,K/yr) is similar to that of FG\,Sge ($350$\,K/yr, \citealt{Jeffery2006}), the 
only other hitherto known LTP object. FG\,Sge was, however, observed only during the final cooling phase, whereas \sao is now the only known 
LTP object that was observed during its blue- as well as red-wards evolution. 
In the upper panel of Fig.~\ref{fig:LTP} we compare the evolution of \sao in the $\log$ \Teff\ -- \logg\ plane with five LTP 
evolutionary tracks that were calculated  from the sequences presented in \cite{MillerBertolami2016} by deliberately removing mass at 
the end of the AGB to obtain LTPs. 
While the 0.50\,\Msol\ track reproduces the low luminosities ($L \propto$ \Teff$^4/g$) of the star, the model predicts a too low 
maximum \Teff\ (only about 28\,kK, hardly enough to ionize the PN). In addition, the evolutionary speed (lower panel of Fig.~\ref{fig:LTP}) 
is too low. Slightly more massive models do reach $\approx 60$\,kK, however, at too low surface gravities. The 0.71\,\Msol\ track matches the 
observed gravities, but heats up to \mbox{\Teff$\approx 100$\,kK} and evolves too quickly. The heating until 1999 is well reflected by 
the 0.53\,\Msol\ model, while the 0.56\,\Msol\ model nicely reproduces the evolutionary speed from 1996 until 2015. 
Fine-tuning of the LTP model parameters may, however, lead to a better agreement of the observations with theoretical 
predictions.\\
Let us now critically examine this LTP scenario in view of other possible evolutionary paths. He-shell flashes may also occur in 
post-EHB stars \citep{Dorman1993} or after the merger of two low-mass stars \citep{zhangetal2012a}. The post-EHB scenario was 
already ruled out in \pa, since it is not possible to explain the young PN. In case of a merger event, He-shell 
flashes occur on too long time-scales ($\approx 20,000$\,yrs, \citealt{zhangetal2012a}) and also the largely solar composition does 
not support a merger scenario. 
Finally, let us consider \sao as a post-RGB star which ended a common-envelope phase (during which the PN was ejected) 
in thermal non-equilibrium. First, we could not find any hint of relative motion of interstellar and photospheric lines by 
comparing the HST/COS observations with the FUSE observations in 2002 and 2006. Therefore, there is no proof for a close companion, 
which would be required for such a scenario. Second, and more importantly, in such a scenario the star would directly, although 
rapidly, evolve to become a white dwarf \citep{DeinzerSengbusch1970, Hall2013}. Therefore \sao would be expected to contract 
further, and not, as shown in this work, expand again. The only way to explain the expansion nevertheless would be, if the star had 
undergone a H-shell flash after entering the white dwarf cooling sequence. Such H-shell flashes occur, however, only when the star 
had reached $L \approx 1$\,\Lsol\ and additionally the cooling rate would be orders of magnitude lower 
\citep{Istrateetal2016, Istrateetal2014, Althausetal2001, Althausetal2013} than what is observed for \sao.
\vspace*{-5mm}
\section{Conclusions}
\label{sect:conclusions}

In this work we have shown, that after the initial rapid heating and contraction (\pa), \sao has 
cooled significantly since 2002 and is now expanding. This can only be explained with a LTP scenario, and indicates that 
the star is now on its way back to the AGB. We stress that \sao is the only LTP object, that was observed during its
blue- and red-wards motion through the Hertzsprung-Russell diagram. The evolutionary speed suggests a central star mass 
between 0.53 and 0.56\,\Msol. However, none of the current LTP models can fully reproduce the evolution of all surface 
parameters simultaneously. In particular the problem with the apparently high gravity of the star should be pursued further. 
A high S/N near-UV spectrum would therefore be highly desirable in order to derive \logg\ more precisely. Refined LTP 
evolutionary calculations, on the other hand, may not only help to explain the nature of \sao, but could 
also provide a deeper insight into the evolution of central stars of PNe as well as the formation of H-deficient stars.
\vspace*{-5mm}
\section*{Acknowledgements}
NR is supported by a Leverhulme Trust Research Project Grant. NR was and TR is supported by the German Aerospace Center 
(DLR, grant 50\,OR\,1507). M3B is partially supported by ANPCyT through grant PICT-2014-2708 and by a Return Fellowship 
from the Alexander von Humboldt Foundation. We thank P. Chayer for reporting to us their results before publication and 
A. Istrate and M. Parthasarathy for helpful discussions and comments.
Based on observations made with the NASA/ESA Hubble Space Telescope, obtained 2015$-$08$-$09 at the Space 
Telescope Science Institute, which is operated by the Association of Universities for Research in Astronomy, Inc., 
under NASA contract NAS 5-26555. These observations are associated with program 13708.
The TMAD service (\url{http://astro-uni-tuebingen.de/~TMAD}) used to compile atomic data and the TIRO service 
(\url{http://astro-uni-tuebingen.de/~TIRO}) used to calculate opacities for this paper were constructed 
as part of the activities of the German Astrophysical Virtual Observatory.
This research has made use of NASA's Astrophysics Data System, the VizieR catalogue access tool and the SIMBAD 
data base operated at CDS, Strasbourg, France.
\vspace*{-5mm}




\bibliographystyle{mnras}
\bibliography{AA}



\appendix

\section{Images}
In the following we compare our best fit TMAP model to the HST/COS observations.

\begin{landscape}
\addtolength{\textwidth}{8.3cm} 
\begin{figure*}
 \centering
\includegraphics[width=220mm]{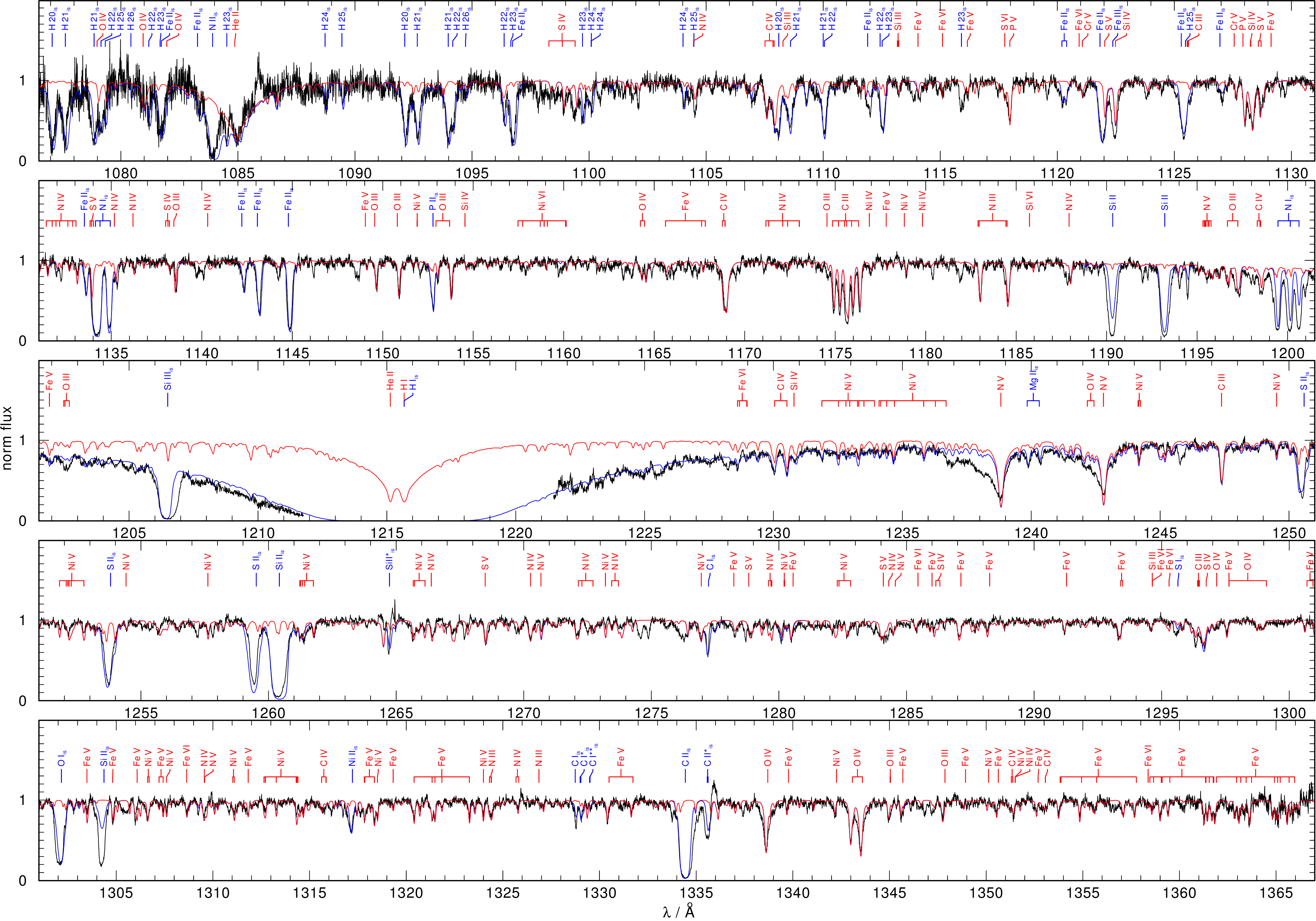}
\caption{G130M HST/COS observation taken in 2015 overplotted with our best fit TMAP model (\Teff = $50$\,kK, \logg =5.5). The red line indicates the pure stellar model spectrum, the blue line additionally includes the interstellar lines. Photospheric and interstellar lines are marked.}
\label{fig:G140M}
\end{figure*}
\end{landscape}

\begin{landscape}
\addtolength{\textwidth}{8.3cm} 
\begin{figure*}
 \centering
\includegraphics[width=220mm]{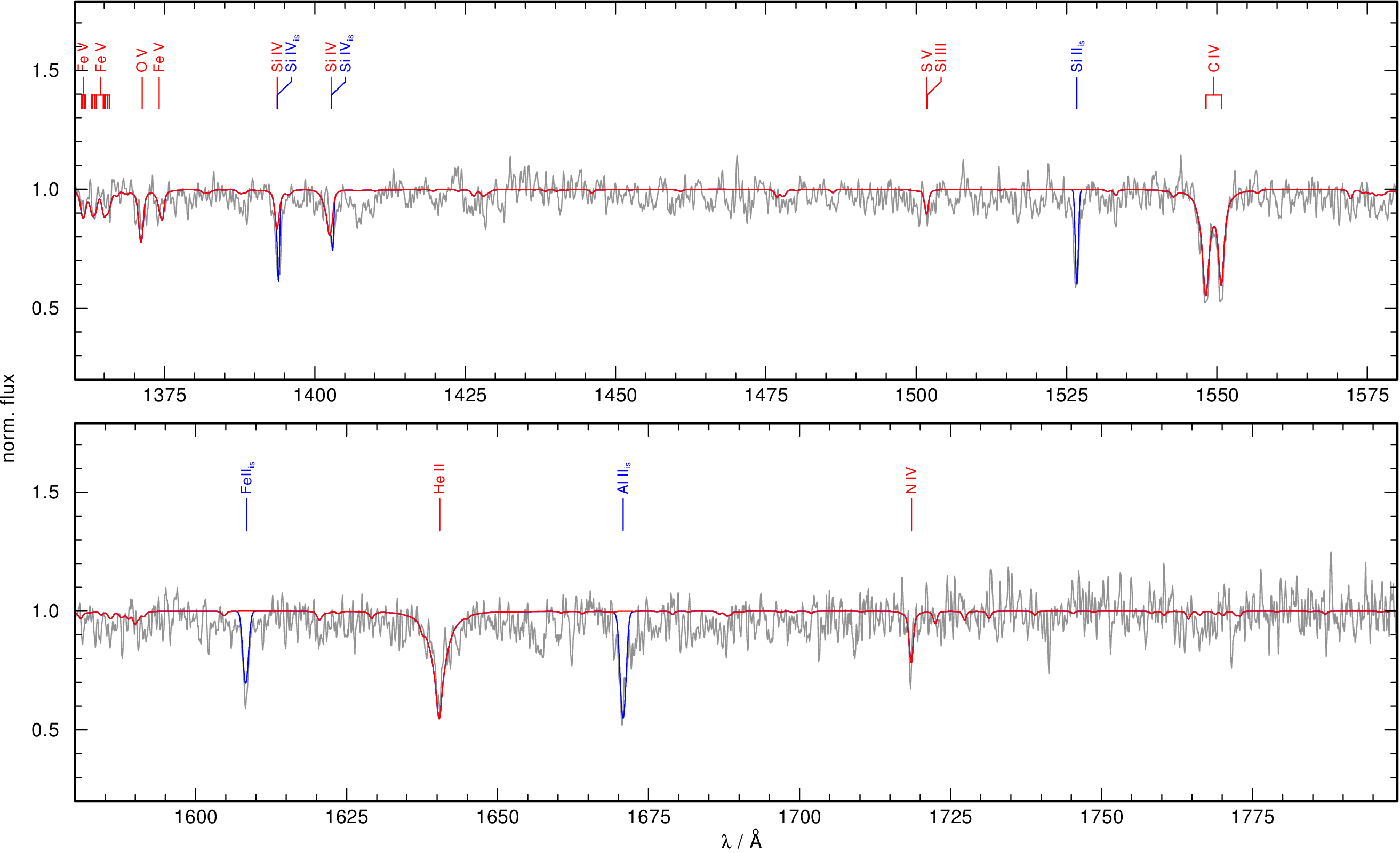}
\caption{Like Fig.~\ref{fig:G140M} for the G140L spectrum.}
\label{fig:G140L}
\end{figure*}
\end{landscape}




\bsp	
\label{lastpage}
\end{document}